\documentclass[aps,pre,twocolumn,superscriptaddress]{revtex4-1}

\usepackage{graphicx}% Include figure files
\usepackage{dcolumn}% Align table columns on decimal point
\usepackage{bm}% bold math
\usepackage{epsf}
\usepackage{amsmath}
\usepackage[colorlinks=true,linkcolor=blue,urlcolor=blue,citecolor=blue,pdfusetitle]{hyperref}
\usepackage{amssymb}                                                                                                                                                                                                                                                                                                                                                                                                                     
\usepackage{xcolor}
\usepackage{array}  
\usepackage{latexsym}
\usepackage{amsfonts}   

\newcommand{\bra}[1]{\langle #1|}
\newcommand{\ket}[1]{|#1\rangle}
\newcommand{\braket}[2]{\left\langle #1|#2\right\rangle}
\newcommand{\tr}[1]{\mathrm{tr}\left\{#1\right\}}

\newcommand{\la}{\left\langle}
\newcommand{\ra}{\right\rangle}

\newcommand{\bla}{bla\\bla\\bla\\bla\\bla}

\newcommand{\PRA}{Phys. Rev. A }

\newcommand{\PRE}{Phys. Rev. E }
\newcommand{\PRX}{Phys. Rev. X }
\newcommand{\PRL}{Phys. Rev. Lett. }

\newcommand{\NJP}{New. J. Phys. }
\newcommand{\NC}{Nature Comm. }
\definecolor{blueOA}{rgb}{0.10, 0.5, 0.81}

\begin{document}

\title{%Assisted adiabatic quantum refrigerator \\ or \\ 
%A superadiabatic quantum Otto refrigerator \\ or \\
Shortcut-to-adiabaticity quantum Otto refrigerator}

\author{Obinna Abah}
%\email{abahobinna@gmail.com}
\affiliation{Centre for Theoretical Atomic, Molecular, and Optical Physics, School of
Mathematics and Physics, Queen's University Belfast, BT7 1NN, United Kingdom}

\author{Mauro Paternostro}
\affiliation{Centre for Theoretical Atomic, Molecular, and Optical Physics, School of
Mathematics and Physics, Queen's University Belfast, BT7 1NN, United Kingdom}

\author{Eric Lutz}

\affiliation{Institute for Theoretical Physics I, University of Stuttgart, D-70550 Stuttgart, Germany}

\begin{abstract}
We investigate the  performance of a quantum Otto refrigerator operating in finite time and exploiting local counterdiabatic techniques. We  evaluate its coefficient of performance and  cooling power  when the working medium consists a quantum harmonic oscillator with a time-dependent frequency. We find that the quantum refrigerator outperforms its conventional counterpart, except for very short cycle times, even when the driving cost of the local counterdiabatic driving is included. We  moreover derive upper   bounds  on the performance of the thermal machine  based  on quantum speed limits and show that they are tighter than the second law of thermodynamics.% {\bf which is sharper than the second law of thermodynamics bounds}. 
\end{abstract}

\date{\today}

\maketitle

%\section{Introduction}
%Cooling is essential for numerous quantum technological applications. 
%Recently, the design and potential realization of quantum thermal machines has attracted considerable attention due to the fundamental need to understand the interplay between thermodynamics and quantum mechanics \cite{gem04}, and the strong technological drive embodied by the research for quantum-enhanced thermodynamic devices. Progress in this field would have profound applications in the quantum technology advancements, for instance error correction in quantum information processing  and efficient cooling \cite{kos13}.

Heat engines and refrigerators are two prime examples of thermal machines.  While heat engines produce  work by transferring heat from a hot to a cold  reservoir, refrigerators consume work to extract heat from a cold to a hot reservoir \cite{cal85,cen01}. Refrigerators thus appear as heat engines functioning in reverse. According to the second law of thermodynamics, the coefficient of performance (COP) of any refrigerator, defined as the ratio of heat input and work input, is limited by the Carnot expression,
$\epsilon_\mathrm{C} ={T_1}/{(T_2 - T_1)}$, where $T_{1,2}$ denote the respective temperatures of the cold and the hot reservoirs \cite{cal85,cen01}.
However, this maximum coefficient of performance is only attainable  in the limit of infinitely long refrigerator cycles where the cooling power vanishes.  At the same time, any refrigerator  that runs in finite time necessarily dissipates irreversible entropy, which reduces its coefficient of performance. 
An important issue is  to optimize the finite-time performance of  thermal machines \cite{and11}. A central  result of the theory of finite-time thermodynamics is  the coefficient of performance  at maximum figure of merit, $\epsilon^\ast = 1/\sqrt{1 - T_1/T_2} - 1$, which is the counterpart of the Curzon-Ahlborn efficiency for heat engines~\cite{yan90,vel97,san04,aba16}. 

Techniques based on {shortcuts to adiabaticity} (STA)~\cite{che10b,tor13}  have recently been suggested as promising candidates to approach such a desired regime of performance-optimized finite-time quantum thermal machines. The implementation of STA methods on an evolving system mimics its adiabatic dynamics in a finite time~\cite{dem03,ber09,che10,mas10,cam13,tor13,def14,cam15,oku17,sch11,bas12,wal12,du16,an16,ste17}. Among the STA techniques put forward so far is the local counterdiabatic (LCD) driving technique, which cancels the possible nonadiabatic transitions induced by the dynamics of a given system by introducing an auxiliary local control potential~\cite{cam13}.  It offers a wide range of applicability and has been experimentally successfully realized in state-of-art ion trap setups \cite{sch11,bas12}. Such STA strategies hold the potential to enhance the performance of both classical and  quantum heat engines~\cite{den13,tu14,cam14,bea16,cho16,aba17,aba17a}.  However, these studies have so far focussed on the unattainability of the absolute zero temperature (according to the third law of thermodynamics)
 in a quantum refrigerator context. Therefore, a full fledged application of STA techniques to enhance the overall performance of a quantum refrigerator is still missing. Moreover, the implementation of STA protocols is not without an energetic cost, which is induced by the additional control potentials. In this regard, only recently the cost of performing STA drivings has been properly taken into account in the performance analysis of quantum heat engines \cite{aba17,aba17a,zhe16,cou16,li18,cak19,tob19}.

In this paper, we study the  STA quantum Otto refrigerator taking into account the cost of the driving in the performance analysis. We explicitly evaluate the coefficient of performance and cooling power of such refrigerator. We find that its performance can exceed its conventional counterpart even when the cost of the STA driving is included, except for very short cycle times. We further use the concept of quantum speed limits for driven unitary dynamics~\cite{DeffnerCampbellReview} to derive generic upper bounds on both the coefficient of performance and the cooling rate of the superadiabatic refrigerator. 

The remainder of this paper is organized as follows. In Sec.~\ref{Otto} we introduce the quantum Otto refrigerator and illustrate the formalism and notation in use in the rest of the article. Section~\ref{super} is dedicated to the analysis of the quantum refrigerator under local counterdiabatic STA driving with Sec.~\ref{performance} discussing its performance. Section~\ref{bounds} is further dedicated to the establishment of  upper bounds on such a performance of the refrigerator as set by the use of the quantum speed limit valid for the dynamics that we explore here. Finally, in Sec.~\ref{conc} we draw our conclusions and discuss the possibility for further developments opened by our assessment.

\section{Quantum Otto refrigerator}
\label{Otto}
\begin{figure}[!t]
\includegraphics[width=0.85\columnwidth]{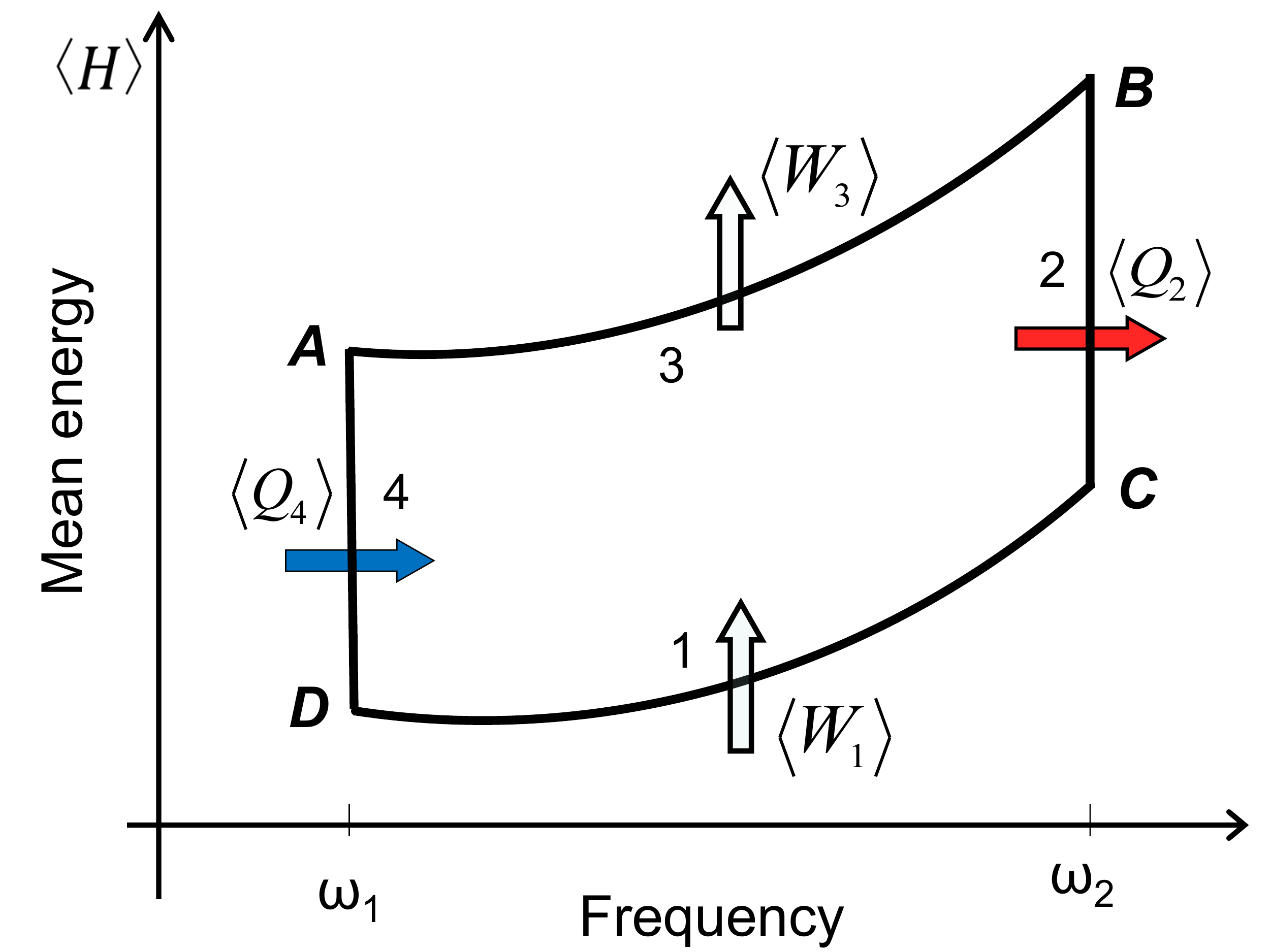}
\caption{Energy-frequency diagram of a generic quantum Otto refrigerator. The thermodynamic cycle consists of two isentropic  processes (strokes 1 and 3) and  two isochoric  processes (strokes 2 and 3). During one complete cycle, work $\la W_1\ra + \la W_3\ra$ is consumed by the quantum refrigerator to pump heat $\la Q_4\ra$ from the cold to the hot reservoir.
\label{cycle}}
\end{figure}
The quantum Otto cycle is a paradigm for thermodynamic quantum devices~\cite{cam14,lin03a,san04,bir08,rez09,kos10,all10,tom12,aba12,yua14,aba16,kar16,rez17,aba19}. The cycle consists of two isentropic and two isochoric processes. At the end of a cycle, work is consumed by the refrigerator to pump heat from a cold to a hot reservoir. In this paper we make the choice of a working medium  embodied by a quantum harmonic oscillator with controllable time-dependent frequency $\omega_t$ (see Fig.~\ref{cycle})  and corresponding Hamiltonian
\begin{equation}
H_\mathrm{O}(t) = \frac{1}{2m}p^2 +  \frac12m\omega_t^2 x^2. 
\label{HOeq}
\end{equation}
Here $m$ is the mass of the oscillator while $x$ ($p$) is its position (momentum) operator. The device is alternately coupled to two heat baths at inverse temperatures $\beta_i = 1/(k_\mathrm{B}T_i) \,  (i\!=\!1, 2)$, where $k_\mathrm{B}$ is the Boltzmann constant. 
Concretely, the Otto cycle consists of the following  four steps as shown in Fig. 1:
\begin{enumerate}
\item \textit{Isentropic compression}, corresponding to the transformation A$(\omega_1,\beta_1)\rightarrow B(\omega_2,\beta_1)$ in Fig.~\ref{cycle}. The frequency is varied during time $\tau_1$, while the system is isolated from the baths. The corresponding evolution is unitary and the von Neumann entropy of the oscillator is constant.
\item \textit{Hot isochore}, associated with the transformation B$(\omega_2,\beta_1)\rightarrow C(\omega_2,\beta_2)$ in Fig.~\ref{cycle}. In this process, the oscillator is weakly coupled to the reservoir at inverse temperature $\beta_2$ at fixed frequency and  
for a time $\tau_2$. Notice that no request is made for thermalization of the oscillator.
\item \textit{Isentropic expansion}, described by the transformation $C(\omega_2,\beta_2)\rightarrow D(\omega_1,\beta_2)$ in Fig.~\ref{cycle}. The frequency of the working medium is unitarily changed back to its initial value during time $\tau_3$. No change of entropy occurs during this stroke.
\item \textit{Cold isochore}, at constant frequency, illustrated by the $D(\omega_1,\beta_2)\rightarrow A(\omega_1,\beta_1)$ process in Fig.~\ref{cycle}. This transformation is obtained by weakly coupling the oscillator to the reservoir at inverse temperature $\beta_1>\beta_2$  and letting the relaxation to the initial thermal state $A(\omega_1,\beta_1)$ occur within a (in general short) time $\tau_4$.
\end{enumerate}
The total cycle time is $\tau_\text{cycle} =\sum^4_{j=1} \tau_j$. In the rest of our analysis, we assume, as commonly done~\cite{lin03a,rez09,tom12,aba16}, that the time needed to accomplish the isochoric transformations is negligible with respect to the compression or expansion times, so that the total cycle time can be approximated to $\tau_\text{cycle} \simeq \tau_1 + \tau_3 = 2\tau$ for equal stroke duration.  This assumption does not affect the generality of our results.

During the first and third strokes (compression and expansion), the quantum oscillator is isolated and only work is performed by changing the frequency in time.  The mean work of the unitary dynamics can be evaluated by using the exact solution of the Schr\"odinger equation for the parametric oscillator for any given frequency modulation~\cite{def08,def10}. The mean work under scrutiny is thus given by \cite{aba16}
\begin{equation}
\begin{aligned}
\la W_1\ra &= \frac{\hbar \omega_2}{2}\left(Q^\ast_1 - \frac{\omega_1}{\omega_2}\right)\coth\left(\frac{\beta_1\hbar\omega_1}{2}\right),\\
\la W_3\ra &=  \frac{\hbar\omega_1}{2}\left(Q^\ast_3 - \frac{\omega_2}{\omega_1}\right)\coth\left(\frac{\beta_2\hbar\omega_2}{2}\right).
\end{aligned}
\end{equation}
We have introduced  the dimensionless quantities $Q^\ast_{1,3}$ that, by depending on the speed of the frequency driving~\cite{hus53}, embodies a parameter of adiabaticity of the dynamics. In general, we have $Q^\ast_{1,3} \ge 1$, with the equality being satisfied for a quasi-static frequency modulation. Its expression is not crucial for the present analysis and can be found in Refs.~\cite{def08,def10}, to which we refer for more details. 

During the thermalization steps (isochoric processes), heat is exchanged with the reservoirs. Such contributions can be quantified by calculating the corresponding variation of energy of the oscillator, which gives us 
%The heat exchanged during the second and fourth  (the cooling part) stroke, $\la Q_4\ra$, reads \cite{aba16},
\begin{equation}
\begin{aligned}
\la Q_2\ra &= \frac{\hbar \omega_2}{2} \left[\coth\left(\frac{\beta_2\hbar\omega_2}{2}\right) - Q^\ast_1\coth\left(\frac{\beta_1\hbar\omega_1}{2}\right)\right],\\
\la Q_4\ra &= \frac{\hbar \omega_1}{2} \left[\coth\left(\frac{\beta_1\hbar\omega_1}{2}\right) - Q^\ast_3\coth\left(\frac{\beta_2\hbar\omega_2}{2}\right)\right].
\end{aligned}
\label{eq3}
\end{equation}
In order to operate as a refrigerator, the system should absorb heat from the cold reservoir, so that $\la Q_4\ra \ge 0$, and release it into the hot reservoir, which entails $\la Q_2\ra \le 0$. According to Eq.~(\ref{eq3}), the condition for cooling is thus that $\omega_2/\omega_1 > \beta_1/\beta_2$. 

The coefficient of performance $\epsilon$ of the quantum Otto refrigerator is given by the ratio of the heat removed from the cold reservoir to the total amount of work performed per cycle, $\epsilon = {\la Q_4\ra}/({\la W_1\ra + \la W_3\ra})$. It explicitly reads~\cite{aba16}
\begin{equation}
\label{epsilon}
\epsilon = \frac{\omega_1[\text{c}(x_1) - Q^\ast_2 \text{c}(x_2)]}{(\omega_2 Q^\ast_1 - \omega_1) \text{c}(x_1) - (\omega_2 - \omega_1Q^\ast_2) \text{c}(x_2)},
\end{equation}
where we have defined $x_j = \beta_j\hbar\omega_j/2~(j=1,2)$ and the function $\text{c}(x_{1,2}) = \coth(x_{1,2})$. For slow (adiabatic) driving processes, $Q^\ast_{1,2} = 1$, the coefficient of performance of the engine becomes  \cite{aba16}
\begin{equation}
\epsilon_\text{AD} = \frac{\omega_1}{\omega_2 - \omega_1},
\label{epsilonAD}
\end{equation} 
which is positive provided that $\omega_2 > \omega_1$.

An upper bound to the coefficient of performance in Eq.~(\ref{epsilon})  follows from the second law of thermodynamics which states that the total entropy  production of a cyclic thermal device is always positive \cite{cal85,ali79}. Employing the quantum relative entropy of two density operators, $S(\rho_1||\rho_2) = \tr {\rho_1 \ln \rho_1 - \rho_1 \ln \rho_2}>0$, 
the total entropy production for one complete cycle can be written as
\begin{eqnarray}
\label{DeltaS}
\Delta S_{\text{tot}} &\!=\!& S(\rho_A||\rho_B) + S(\rho_B||\rho_C) + S(\rho_C||\rho_D) + S(\rho_D||\rho_A)  \nonumber\\
&=& -\beta_2 \la Q_2\ra - \beta_1 \la Q_4\ra \ge 0,
\end{eqnarray}
where we have used the fact that the quantum relative entropy during the isentropic processes $AB$ and $CD$ is null, as the von Neumann entropy is constant. Moreover, the quantum relative entropy of the isochoric processes $BC$ and $DA$ corresponds to the entropy production associated with the heating and cooling steps. 
From Eq.~\eqref{epsilon}, the total entropy production is then \cite{all10,rez17}
\begin{eqnarray}
\label{DeltaS1}
\Delta S_\text{tot} 
&\!=\!&x_2[Q^\ast_1 \text{c}(x_1) - \text{c}(x_2)] - x_1[\text{c}(x_1) - Q^\ast_3 \text{c}(x_2)] \ge 0. 
\label{entropy}\hspace{0.5cm}
\end{eqnarray}
 Equality to zero is reached for the Carnot cycle scenario for which $\beta_2/\beta_1\!=\!\omega_1/\omega_2$.
Based on the first law of thermodynamics, we have in addition
\begin{equation}
\label{8}
-(\la W_1\ra + \la W_3\ra) = \la Q_2\ra + \la Q_4\ra.
\end{equation} 
Combining Eqs.~\eqref{DeltaS} and \eqref{8}, we obtain the following upper bound on the refrigerator performance
\begin{equation}
\frac{\la W_1\ra + \la W_3\ra}{\la Q_4\ra} \le \frac{\beta_2}{\beta_1 - \beta_2} = \frac{T_1}{T_2 - T_1} = \epsilon_\text{C}.
\end{equation}
The above equation shows that the coefficient of performance of the quantum refrigerator is always bounded by the Carnot coefficient of performance.

\section{Driving a quantum refrigerator with shortcuts to adiabaticity}
\label{super}
Let us now consider the situation when the compression and expansion strokes of the  Otto refrigerator cycle is sped up by addition of  a counterdiabatic driving control field $H_\mathrm{STA}^\mathrm{CD}(t)$ to the original harmonic oscillator Hamiltonian $H_\mathrm{O}(t)$. Scope of this term is to suppress the non-adiabatic transitions induced by the finite-time evolution of the oscillator and, as a consequence, quench the entropy production all the way down to the value taken in the adiabatic manifold of the initial system Hamiltonian. 
The resulting effective Hamiltonian  reads \cite{dem03,ber09}
\begin{equation}
\begin{aligned}
H_\text{CD}(t)   &= H_\mathrm{O}(t) +  H_\mathrm{STA}^\mathrm{CD}(t)\\
& = H_\text{O}(t) + i\hbar \sum_n\left(\ket{\partial_t n}\bra{n} - \braket{n}{\partial_t n} \ket{n}\bra{n}\right),
\end{aligned}
\end{equation}
where $\ket{n} \equiv \ket{n(t)}$ denotes the $n^{\rm th}$ eigenstate of the original Hamiltonian $H_\text{O}(t)$. For a harmonic working medium, the counterdiabatic term  $H_\mathrm{STA}^\mathrm{CD}(t)$ is  \cite{che10,tor13}
\begin{equation}
H_\mathrm{STA}^\mathrm{CD}(t) = -\frac{\dot{\omega}_t}{4 \omega_t} (x p + p x).
\label{Hstacd}
\end{equation}
Although this additional control removes the requirement of slow driving, the (non-local) counterdiabatic potential -- which induces squeezing of the oscillator -- makes its experimental application/implementation a challenging task. As a result, in order to circumvent this difficulty, it is natural to construct a unitarily-equivalent Hamiltonian with a local potential. This is achieved by applying the operator $U_x = \exp\left({i m \dot{\omega}_t x^2}/{4 \hbar \omega}\right)$, which cancels the squeezing term and gives the new effective local counterdiabatic (LCD) Hamiltonian~\cite{cam13}
\begin{equation}
H_\mathrm{LCD}(t) = U_x^\dagger (H_\mathrm{CD}(t) - i\hbar \dot{U}_x U_x^\dagger) U_x = \frac{p^2}{2 m} + \frac{m\Omega_t^2 x^2}{2},
\label{19}
\end{equation}
where the modified time-dependent frequency is $\Omega^2(t) = \omega_t^2 -{3\dot{\omega}_t^2}/{4\omega_t^2} + {\ddot{\omega}_t}/{2\omega_t}$. By requesting that the initial and final state of the working medium ensuing from $H_\mathrm{LCD}(t)$ equal that from the original Hamiltonian $ H_\mathrm{O} (t)$, one gets the boundary conditions
\begin{equation}
\begin{array}{lcr}
\omega_0= \omega_i,& \dot{\omega}_0 = 0 ,& \ddot{\omega}_0 = 0,\\
\omega_\tau =\omega_f,& \dot{\omega}_\tau = 0 ,& \ddot{\omega}_\tau = 0,
\label{BC}
\end{array}
\end{equation}
where $\omega_{i,f} = \omega_{1,2}$ correspond to the initial and final frequency of the compression/expansion strokes. 
A suitable ansatz is \cite{cam13,def14}\begin{equation}
\omega_t= \omega_i + 10(\omega_f - \omega_i)s^3 - 15(\omega_f - \omega_i)s^4 +  6(\omega_f  - \omega_i) s^5
\end{equation}
with $s = t/\tau$.
In order to ensure that the trap is not inverted, one must also guarantee that $\Omega(t)^2 > 0$ is always fulfilled \cite{cam14}. 
The mean value of the local counterdiabatic Hamiltonian $H_\mathrm{LCD}(t)$ may be calculated explicitly for an initial thermal state  and reads \cite{aba17}
\begin{eqnarray}
\langle H_\mathrm{LCD}(t) \rangle &=& \frac{\hbar \omega_t}{2} \left(1 - \frac{\dot{\omega}_t^2}{4 \omega_t^4} + \frac{\ddot{\omega}_t}{4 \omega_t^3} \right)\coth\left(\frac{\beta\hbar\omega_i}{2} \right) ,\nonumber\\
 &=& \frac{\omega_t}{\omega_0} Q_\mathrm{LCD}^\ast \langle H(0)\rangle,
\end{eqnarray} 
where we have introduced the LCD parameter
\begin{equation}
Q^\ast_\text{LCD} (t) = 1 - \frac{\dot{\omega}_t^2}{4 \omega_t^4} + \frac{\ddot{\omega}_t}{4\omega_t^3}.
\end{equation}
The expectation value of the control field $H_\mathrm{STA}^\mathrm{LCD}(t) $ follows therefore as
\begin{equation}
\la H_\mathrm{STA}^\mathrm{LCD}(t) \ra 
=\frac{\hbar \omega_t}{2} \left( -\frac{\dot{\omega}_t^2}{4 \omega_t^4} + \frac{\ddot{\omega}_t}{4 \omega_t^3}\right) \coth\left(\frac{\beta\hbar\omega_i}{2} \right), \hspace{0.5cm}
\end{equation}
where we have used $\la H_\mathrm{O}(t)\ra\!=\!\hbar\omega_t\coth(\beta\hbar\omega_i/2)/2$ \cite{def10}. Based on the boundary conditions in Eq.~\eqref{BC}, we have $\la H_\mathrm{STA}^\mathrm{LCD}(\overline{t})\ra \!=\!0$ for $\overline{t}=0$ and $\tau$, while the time-averaged value is non-null. We  also remark that the local counterdiabatic control has been implemented in various experimental platforms~\cite{sch11,wal12}, specifically in  Paul traps \cite{wal12} which are a potential candidate for building quantum thermal devices~\cite{ros16}. %\blueComm{problems with references!}

\begin{figure}[!t]
%\begin{minipage}{\columnwidth}
%\includegraphics[width=0.75\columnwidth]{fig2}
%\end{minipage}
%\begin{minipage}{\columnwidth}
\includegraphics[width=0.8\columnwidth]{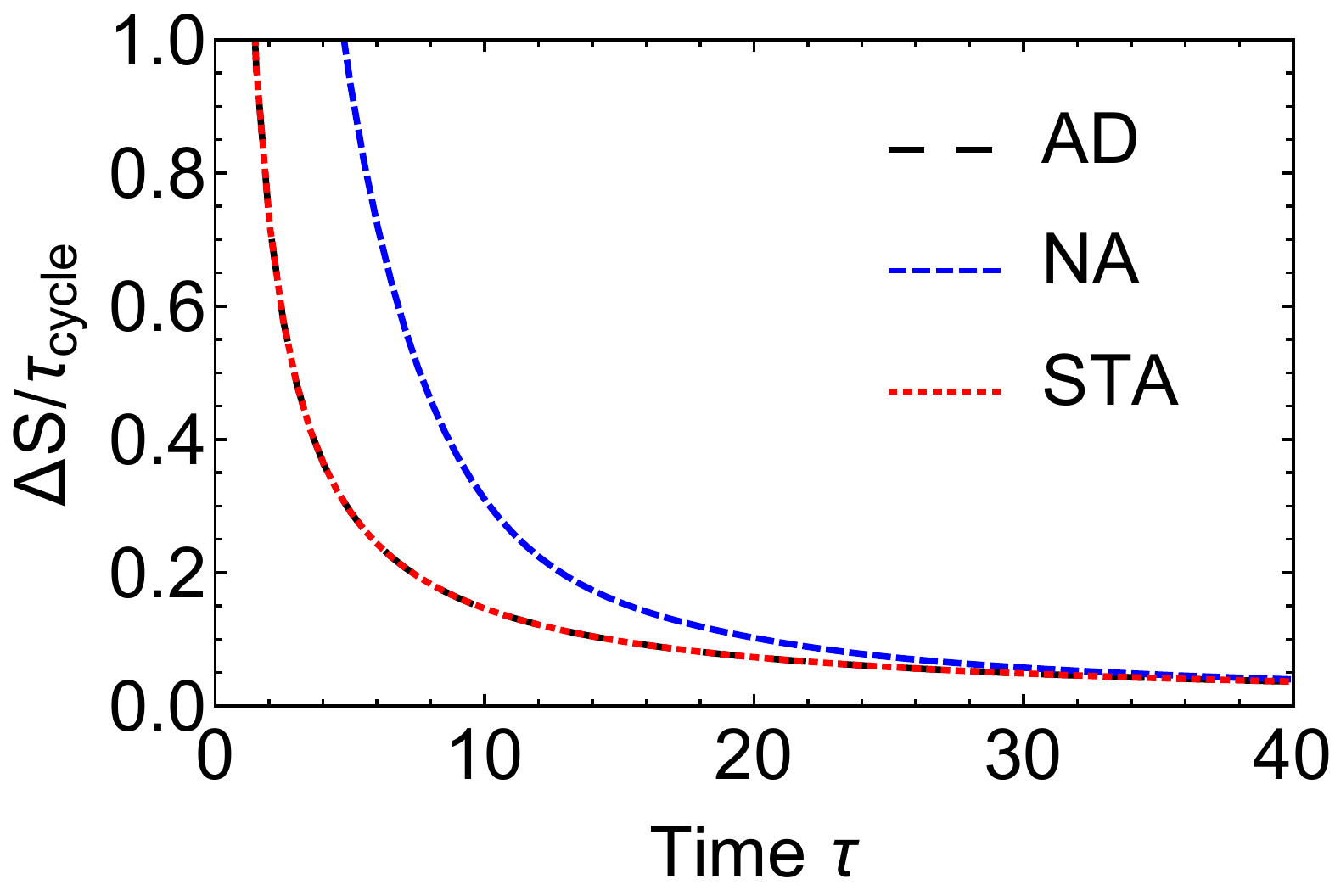}
%\end{minipage}
\caption{Entropy production rate $\Delta S_\text{tot}/\tau_\text{cycle}$  of the quantum Otto refrigerator plotted against the driving time $\tau$. The blue (small dashed) line shows the nonadiabatic expression Eq.~\eqref{entropy} in the absence of STA driving, while the red (dotted) line represents the corresponding result including STA techniques. Local counterdiabatic driving is seen to greatly reduce the irreversible entropy production rate to the adiabatic value (black large dashed). Parameters  are $\hbar = 1, \omega_1 = 0.1$, $\omega_2 = 0.5$, $\beta_1 = 1$ and $\beta_2 = 0.75$. %\textit{Should we add the rate of entropy production plot that go to zero in long time? - the lower plot}
}
\label{fig2}
\end{figure}

Figure~\ref{fig2} shows  the rate of entropy production $\Delta S_\text{tot}/\tau_\text{cycle}$ as a function of  the  time $\tau$ for adiabatic and nonadiabatic driving. We see that for short cycle time, the entropy production of nonadiabatic transition processes dramatically increases (blue dashed), thus leading to lower performance of the thermal machine. On the other hand, the application of STA methods is effective in suppressing such over-shooting of irreversible entropy  (red dotted) to the adiabatic value (black large dashed).

\section{Performance of a superadiabatic quantum refrigerator}
\label{performance}
We now study three important quantities characterizing the performance of a refrigerator, namely its   coefficient of performance $\varepsilon$,  cooling rate $J_\mathrm{STA}^c$ and  figure of merit $\chi$. Taking into account the energetic cost of the STA driving, we define the coefficient of performance of  the superadiabatic quantum Otto refrigerator  as the ratio of the heat removed from the cold reservoir to the total amount of energy added per cycle
\begin{equation}
\label{epsilonSA}
\epsilon_\mathrm{STA} %&=& \frac{\mathrm{Energy\, output/benefit}}{\mathrm{Energy\, input/cost}}\nonumber\\
=
 \frac{\la Q_4\ra}{\sum_{i=1,3}\left(\la W_i\ra_\mathrm{STA} + \la H^i_\mathrm{STA}\ra_\tau \right)} ,
\end{equation}
where  $\la H^i_\text{STA}\ra_\tau = (1/\tau) \int_0^\tau dt \la H^i_\text{STA}(t)\ra$  $(i=1,3)$, is the time-average of the mean value  of the local potential for the compression/expansion strokes and quantifies 
the energetic cost of the transitionless driving. When the energetic cost of the STA protocol is ignored (which corresponds to setting $\la H^i_\text{STA}\ra_\tau =0$ in Eq.~(\ref{epsilonSA})), the coefficient of performance reduces to the adiabatic expression  $\epsilon_\mathrm{AD}$ given by Eq.~(\ref{epsilonAD}) \cite{rez09,bir08,yua14,aba16}.

\begin{figure}[!]
\includegraphics[width=0.8\columnwidth]{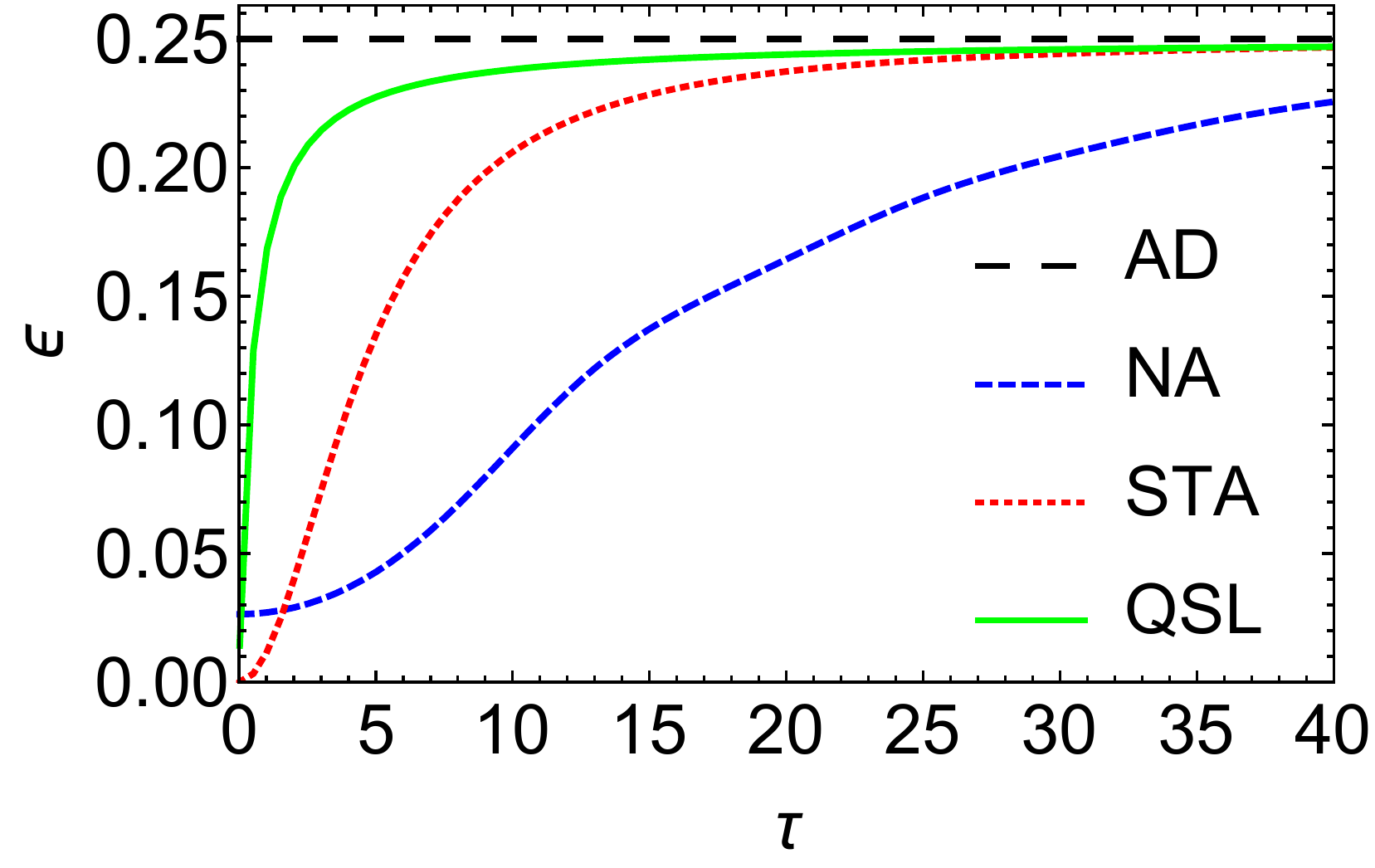}
\caption{Coefficient of performance of the quantum Otto refrigerator as a function of time $\tau$. The blue (small dashed) line shows the exact nonadiabatic case (NA), Eq.~(\ref{epsilon}), while the red (dotted) line and the green (solid) lines respectively display the STA results, Eq.~(\ref{epsilonSA}) and the quantum speed limit (QSL) bound, Eq.~(\ref{epsilonqsl}). The black (large dashed) line corresponds to the adiabatic case, Eq.~(\ref{epsilonAD}). Same parameters as in Fig.~\ref{fig2}.
}
\label{3}
\end{figure}

Figure~\ref{3} shows the coefficient of performance  of the superadiabatic quantum refrigerator $\epsilon_\mathrm{STA}$ (red dotted) as a function of  time $\tau$, together with the adiabatic  $\epsilon_\mathrm{AD}$ (black large dashed) and nonadiabatic $\epsilon_\text{NA} = \la Q_4\ra/(\la W_1\ra+\la W_3\ra)$ (blue small dashed) counterparts.  We observe  that the superadiabatic driving significantly enhances the performance of the quantum Otto refrigerator, $\epsilon_\mathrm{NA} \leq \epsilon_\mathrm{STA} \leq  \epsilon_\mathrm{AD}$, for all driving times larger than $\tau\approx 2.0$, even though the energetic cost of the STA is explicitly included. We additionally note that the superadiabatic coefficient of performance $\epsilon_\mathrm{STA}$  is remarkably close to the adiabatic value $\epsilon_\mathrm{AD}$ for  $\tau\geq 25$, indicating that the energetic STA cost is relatively small for larger times. Yet, the nonadiabatic coefficient of performance $\epsilon_\mathrm{NA}$ is already greatly reduced compared to the adiabatic value in this regime. The STA techniques thus appear here to be highly effective in suppressing nonadiabatic transitions at a  little cost. 

\begin{figure*}[!]
%{\bf (a)}\hskip5.5cm{\bf (b)}\hskip5.5cm{\bf (c)}\\
\includegraphics[width=0.69\columnwidth]{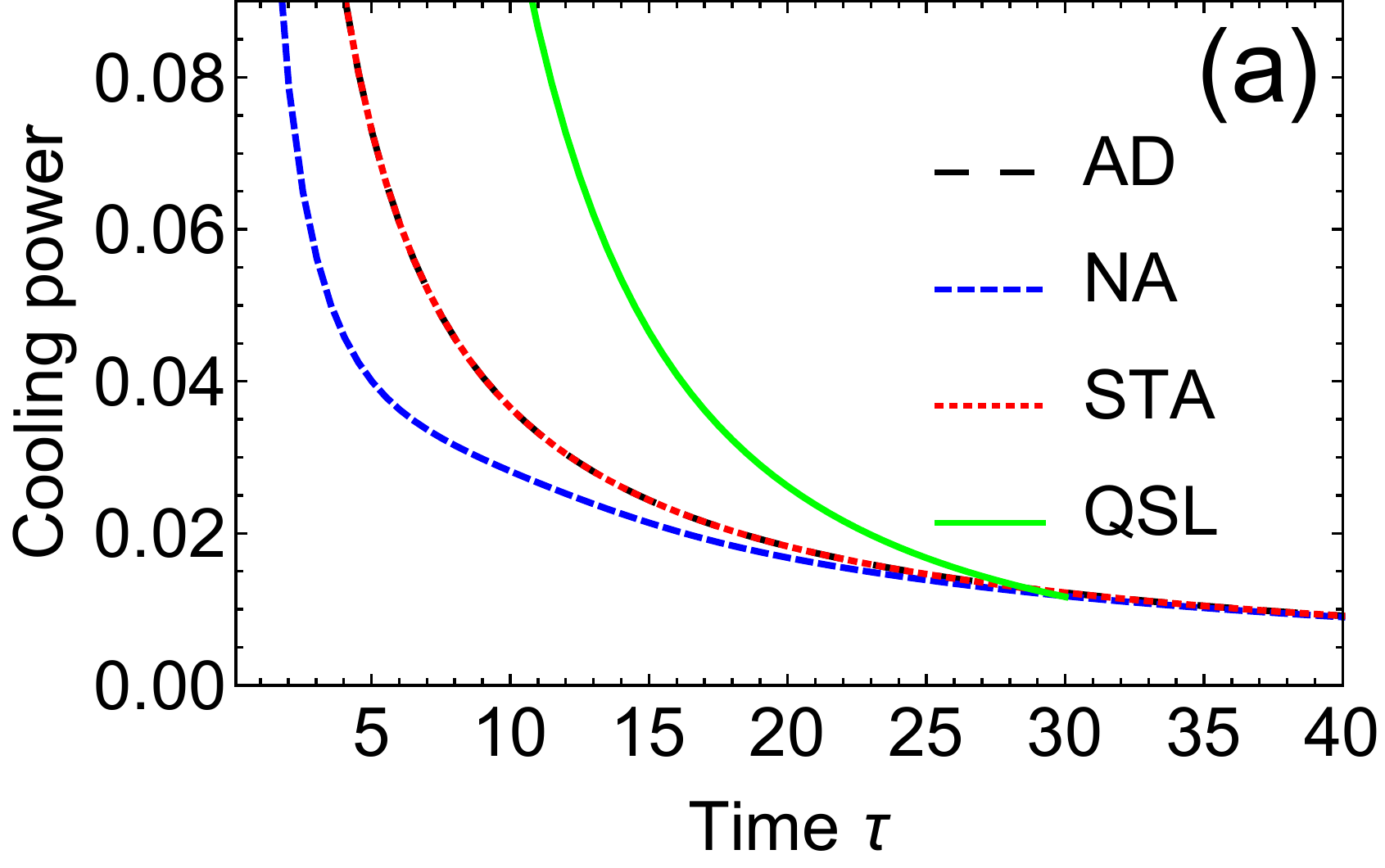}\hfil
\includegraphics[width=0.68\columnwidth]{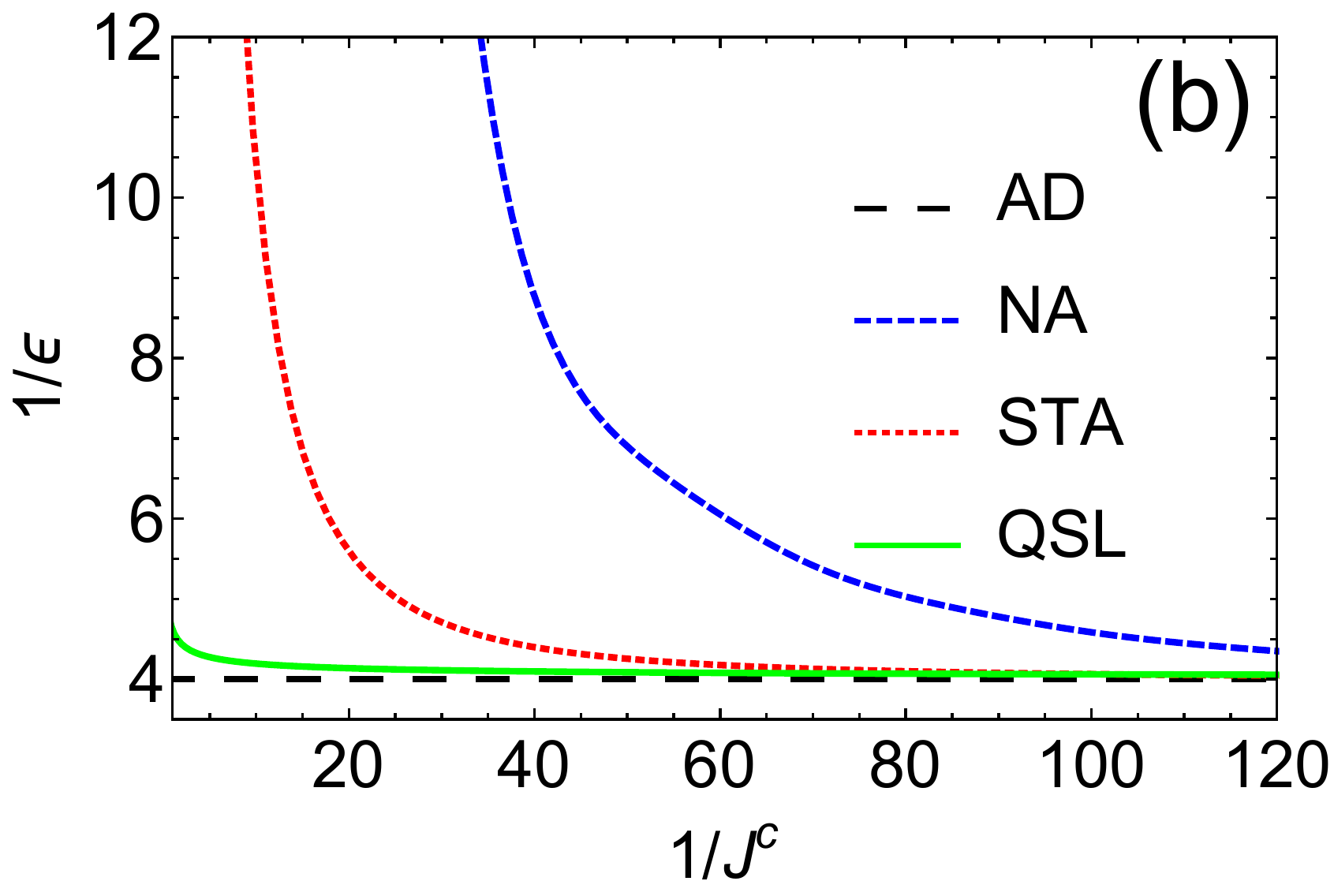}\hfil
\includegraphics[width=0.69\columnwidth]{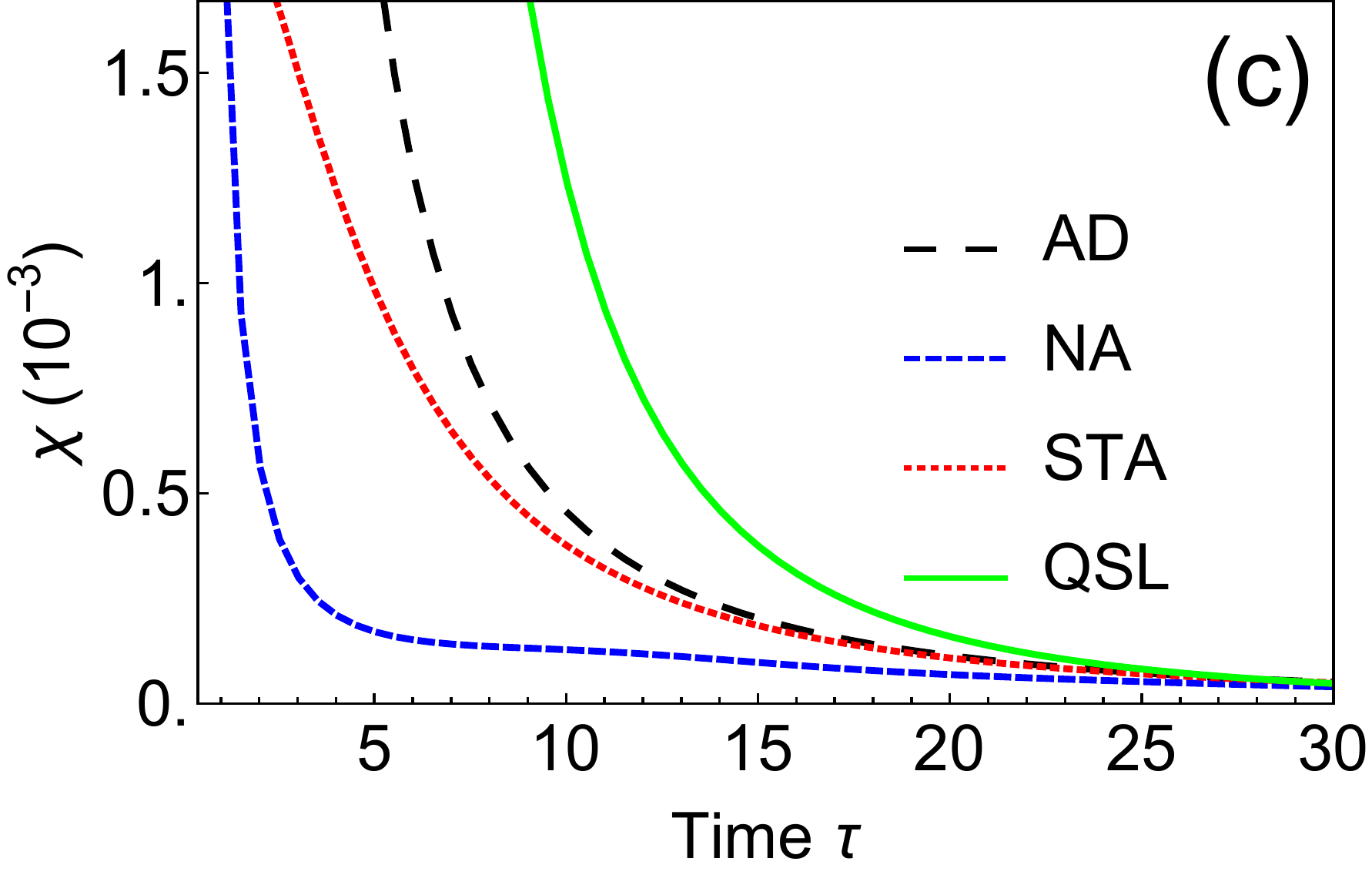}
\caption{(a) Cooling power of the quantum Otto refrigerator as a function of time $\tau$.  The blue (dashed) line shows the exact nonadiabatic case, while the red (dotted) line and the green (solid) lines respectively represent the STA-based result in Eq.~(\ref{JSTA}) and the quantum speed limit bound in Eq.~(\ref{Jqsl}).  The black (large dashed) line is the adiabatic case, Eq.~(\ref{epsilon}).  {(b)} Inverse coefficient of performance as a function of the  cooling power for the time duration $\tau = 0.5 - 45$, for the same cases as in (a).   {(c)} Figure of merit $\chi$ [see Eq.~(\ref{chi_sta})] as a function of the driving time $\tau$ for the same cases as above.  Same parameters as in Fig.~\ref{fig2}.}
\label{fig4}
\end{figure*}

On the other hand, the cooling power of the superadiabatic refrigerator is given by the ratio of heat flowing from the cold reservoir into the system to the cycle time
\begin{equation}
J_\mathrm{STA}^c = \frac{\la Q_4\ra_\mathrm{STA}}{\tau_\mathrm{cycle}}.
\label{JSTA}
\end{equation}
An infinitely long cycle time, which would allow to achieve the maximum coefficient of performance, would thus also gives zero cooling power. In this regard, the main advantage of the STA approach is to realize the same amount of heat output as in the adiabatic case, but in a shorter cycle time. Hence, the STA strategy ensures that $J_\mathrm{STA}^c$ (red dotted) is always greater than the 
nonadiabatic cooling rate $J_\mathrm{NA}^c = \la Q_4\ra_\mathrm{NA}/\tau_\mathrm{cycle}$ (blue dashed) for fast cycles, as shown in  Fig. \ref{fig4}({a}). However, there still exists a trade-off between cooling power and coefficient of performance of STA refrigerator for fast cycles.

Following Feldmann and Kosloff~\cite{fel12}, such trade-off can be illustrated as in Fig.~\ref{fig4}({b}), where the dependence of $1/\epsilon$ on the inverse cooling power $1/J^c$ is illustrated for both the STA driving and the nonadiabatic protocol. The former simultaneously enhances both the coefficient of performance and the cooling power,  thereby clearly demonstrating the benefits of the STA quantum Otto refrigerator over the conventional ones.

We finally consider the figure of merit $\chi\!=\!{\epsilon \la Q_4\ra}/{\tau_\text{cycle}}\!=\!\epsilon J^c$ defined as the product of the coefficient of performance $\epsilon$ and the cooling power of the refrigerator~\cite{vel97,all10,tom12,aba16}. The corresponding expression for a heat engine, $\chi_\text{engine} = \eta \la Q_2\ra/\tau_\text{cycle}\!=\!-\la W\ra/\tau_\text{cycle}$, is equal to its power output $\la Q_2\ra$ being in this case the heat absorbed from the hot reservoir. In optimization problems, the maximum figure of merit (and not the maximum cooling power) condition   for refrigerators is in direct correspondence to the maximum power criterion for heat engines~\cite{vel97,all10,tom12,aba16}.  Figure~\ref{fig4}({c}) presents the corresponding values as a function of $\tau$ for the case of adiabatic, non-adiabatic and STA strategies. In analogy to the cooling power (19), a clear hierarchy emerges as $\chi_\text{NA} \le \chi_\text{STA} \le \chi_\text{AD}$ with the equalities holding in the long-time limit. Compared to the nonadiabatic case, the STA approach increases the area under the curve, which determines  the overall performance of the device.

\section{Performance bounds by quantum speed limit}
\label{bounds}
The maximum performance of a classical thermal machine (refrigerator/engine) is limited by the second law of thermodynamics \cite{cen01}. However, quantum mechanics imposes restrictions on the time of evolution of quantum processes. Understanding such restrictions is important for the  successful implementation of the STA technique~\cite{aba17}.  We next derive general upper bounds for both the STA-based coefficient of performance and cooling rate of the quantum Otto refrigerator using the concept of quantum speed limits, which can be regarded as an extension of the Heisenberg energy-time uncertainty relation \cite{DeffnerCampbellReview,ana90,vai92,uff93,mar98}.

For  unitary driven dynamics, a Margolus-Levitin-type bound \cite{mar98}  on the evolution time given by~\cite{def13a}
\begin{equation}
\tau \geq \tau_\text{QSL}= \frac{\hbar\, \mathcal{L}(\rho_i,\rho_f)}{\la H_\text{STA}\ra_\tau}, 
\label{tqsl}
\end{equation}
is appropriate. Here $\mathcal{L}(\rho_i,\rho_f) = \arccos \sqrt{F(\rho_f,\rho_i)}$ denotes the Bures angle between  initial and final density operators of the system, with  $F(\rho_f,\rho_i)$ the fidelity between the two, and 
$\la H_\text{STA}\ra_\tau$ the time-averaged superadiabatic energy. Eq.~\eqref{tqsl} becomes  a proper bound for the compression and expansion phases, when the engine dynamics is dominated by the STA driving for small $\tau$.

From Eqs.~\eqref{epsilonSA} and  \eqref{tqsl}, an   upper bound on the STA-based coefficient of performance of the quantum Otto refrigerator is obtained as
\begin{equation}
\epsilon_\text{STA} \leq \epsilon_\text{STA}^\text{QSL} = \frac{\la Q_4\ra_\text{AD}}{\la W_{1}\ra_\text{AD} + \la W_{3}\ra_\text{AD} +\hbar(\mathcal{L}_1+\mathcal{L}_3 )/\tau},\label{epsilonqsl}
\end{equation}
where $\mathcal{L}_i$ $(i\!=\!1,3)$ are the respective Bures angles for the compression/expansion steps. Likewise, an upper bound on the STA-based cooling power (Eq.~\ref{JSTA}) reads
\begin{equation}
J^c_\text{STA}\leq J_\text{STA}^{c\text{QSL}} = -\frac{\la Q_4\ra_\text{AD}}{\tau^1_\text{QSL}+\tau^3_\text{QSL}}, \label{Jqsl}
\end{equation}
where $\tau^i_\text{QSL}$ $(i\!=\!1,3)$ are the respective  speed-limit bounds in Eq.~\eqref{tqsl} for the compression/expansion phases. In addition, an upper bound for the figure of merit $\chi$ follows as
\begin{equation}
{\chi_\text{STA} \leq \chi_\text{STA}^\text{QSL} = \epsilon_\text{STA}^\text{QSL} J_\text{STA}^{c\text{QSL}}}.\label{chi_sta}
    \end{equation}
The above upper bounds are displayed in Figs.~\ref{3}, \ref{fig4}({a)}, \ref{fig4}(b) and \ref{fig4}({c)} (green solid). We observe that the quantum bound on the coefficient of performance (Figs. \ref{3} and \ref{fig4}(b))  is much tighter than the adiabatic bound (black large dashed) imposed by the second law of thermodynamics (discussed in Sec.~\ref{Otto}). They are hence more useful for applications. We emphasize that these results are general and do not depend on the choice of the engine cycle or on the STA driving protocol.

\section{Conclusions}
\label{conc}
We have studied the performance of a quantum Otto refrigerator with a working medium consisting of a time-dependent harmonic oscillator, exploiting  STA mechanisms. We have explicitly analyzed the coefficient of performance, the cooling power, as well as the related figure of merit, for the case of local counterdiabatic driving. We have found that the  STA quantum refrigerator  outperforms its conventional nonadiabatic counterpart, except for short  cycle durations, by strongly minimizing the nonequilibrium entropy production, even when the energetic cost of the STA driving is included. We have further derived generic upper bounds  on the coefficient of performance of the Otto refrigerator by using  the concept of quantum speed limits. Such bounds are tighter than those based on the second law of thermodynamics and therefore more useful. The possibility to achieve simultaneous enhancements of coefficient of performance and cooling power should be of advantage for the future design of micro- and nano-devices operating in the quantum regime. %Our findings should be useful in implementing high efficient cooling in quantum regime.

\acknowledgments
We acknowledge support from the Royal Commission for the Exhibition of 1851, the EU Collaborative project TEQ (grant agreement 766900), the DfE-SFI Investigator Programme (grant 15/IA/2864), COST Action CA15220, the Royal Society Newton Fellowhsip (Grant Number NF160966), the Royal Society Wolfson Research Fellowship ((RSWF\textbackslash R3\textbackslash183013), the Leverhulme Trust Research Project Grant (grant nr.~RGP-2018-266) and the DFG (Contract No FOR 2724).

%\newpage

\end{document}